\documentclass[10pt, conference, letterpaper]{IEEEtran}
\IEEEoverridecommandlockouts
\usepackage{cite}
\usepackage{amsmath,amssymb,amsfonts}
\usepackage{algorithmic}
\usepackage{algorithm}
\usepackage{graphicx}
\usepackage{textcomp}
\usepackage{xcolor}
\usepackage{comment}
\usepackage{xspace}
\usepackage{upgreek}
\usepackage{url}
\usepackage{subcaption}
\usepackage{subfig}
\usepackage{multirow}
\newcommand*{\note}[1]{\textcolor{red}{#1}}
\newcommand*{\ye}[1]{\textcolor{black}{#1}}

\newcommand{\algname}{CCRSat\xspace}
\def\BibTeX{{\rm B\kern-.05em{\sc i\kern-.025em b}\kern-.08em
    T\kern-.1667em\lower.7ex\hbox{E}\kern-.125emX}}

\begin{document}


\title{\algname: A Collaborative Computation Reuse Framework for Satellite Edge Computing Networks\\

}


\author{\IEEEauthorblockN{Ye Zhang$^{1}$, Zhishu Shen$^{1}$\textbf{\IEEEauthorrefmark{2}}\thanks{\IEEEauthorrefmark{2} Corresponding author (z\_shen@ieee.org).}, Dawen Jiang$^{1}$, Xiangrui Liu$^{1}$, Qiushi Zheng$^{2}$, and Jiong Jin$^{2}$}

\IEEEauthorblockA{\textsuperscript{$^{1}$}School of Computer Science and Artificial Intelligence, Wuhan University of Technology, China\\
}


\IEEEauthorblockA{\textsuperscript{$^{2}$}School of Science, Computing and Engineering Technologies, Swinburne University of Technology, Australia\\
}
}
\maketitle

\begin{abstract}

In satellite computing applications, such as remote sensing, tasks often involve similar or identical input data, leading to the same processing results. Computation reuse is an emerging paradigm that leverages the execution results of previous tasks to enhance the utilization of computational resources. While this paradigm has been extensively studied in terrestrial networks with abundant computing and caching resources, such as named data networking (NDN), it is essential to develop a framework appropriate for resource-constrained satellite networks, which are expected to have longer task completion time. In this paper, we propose \algname, a collaborative computation reuse framework for satellite edge computing networks. \algname initially implements local computation reuse on an independent satellite, utilizing a satellite reuse status (SRS) to assess the efficiency of computation reuse. \ye{Additionally, an inter-satellite computation reuse algorithm is introduced, which utilizes the collaborative sharing of similarity in previously processed data among multiple satellites.} The evaluation results tested on real-world datasets demonstrate that, compared to comparative scenarios, our proposed \algname can significantly reduce task completion time by up to 62.1\% and computational resource consumption by up to 28.8\%.

\end{abstract}


\section{Introduction}
To meet the evolving connectivity needs of the future digital society in the era of next-generation wireless network technologies beyond the fifth and sixth generations (B5G/6G), satellite networks are the potential solution to provide network coverage with enhanced throughput~\cite{CentenaroCST21,KodheliCST21,Al-HraishawiCST23}. With recent advancements in microelectronics and associated computing technologies, the computing capabilities of modern satellite systems have been significantly improved. Consequently, these systems are now widely utilized across diverse fields, including remote sensing and navigation~\cite{ShenCSUR23}. However, the computing resources on real-world satellites remain relatively limited~\cite{WangIoT23}. Optimizing the utilization of these valuable resources while maintaining satisfactory task processing performance remains both challenging and imperative.

In typical applications such as remote sensing, data processing tasks for similar services with comparable input data tend to result in redundant computation, leading to unnecessary resource consumption. Computation reuse is an emerging paradigm that utilizes existing execution results from previous services or functions for future computations, thereby enhancing resource efficiency~\cite{He2017Exp,NourNW20,Barrios2023Service}. Guo \textit{et al.} introduced FoggyCache that facilitates cross-device approximate computation reuse. Based on FoggyCache, an adaptive locality-sensitive hashing (A-LSH) and homogeneous k-nearest neighbors (H-kNN) algorithm are designed to minimize redundant computations~\cite{guo2018foggycache}. The same authors also designed Potluck~\cite{guo2018potluck} to achieve approximate deduplication across applications. Both FoggyCache and Potluck focus on computation reuse on end-user devices. In the context of edge networks, Mastorakis \textit{et al.} introduced ICedge, a general network framework based on named data networking (NDN), which simplifies service invocation and enhances the reuse of redundant computations at the edge~\cite{mastorakis2020icedge}. Similarly, Al Azad \textit{et al.} proposed Reservoir, a framework that facilitates pervasive computation reuse at edge computing devices while imposing minimal overhead on user devices and the network infrastructure~\cite{al2022reservoir}. In terms of data privacy, Nour \textit{et al.} developed an architecture incorporating federated edge computing to select the appropriate devices with high-quality data to improve data processing performance and reduce communication costs. This architecture also employs computation reuse to handle incoming tasks with minimal computation~\cite{Nour2022Federated}.

Computation reuse provides the benefit of reducing task completion time for services with highly similar task inputs. Therefore, the computation reuse paradigm has already been implemented in real-time applications like intelligent vehicles~\cite{ChenTIV23} and augmented reality~\cite{YushanCST21}. Similarly, in satellite applications such as meteorological monitoring and disaster warning, real-time data processing is essential~\cite{Leyva-MayorgaTC23,GaoIoT24}. Although pre-trained AI models can be deployed on real-world satellites for task processing~\cite{WangIoT23}, handling large-volume tasks, such as high-resolution image processing, remains time-consuming. Therefore, designing a computation reuse solution for resource-constrained satellites is crucial to efficiently handle these computationally intensive tasks. 
Currently, substantial efforts are being made to harness satellite caching and collaborative computing resources to optimize the use of limited satellite computational resources while minimizing processing latency. For example, Hao \textit{et al.} explored the joint optimization of computation offloading, radio resource allocation, and cache placement in low Earth orbit satellite (LEO) networks to minimize the total latency for all ground IoT devices~\cite{Hao2023Joint}. Qiu \textit{et al.} utilized a deep Q-learning (DQN) approach to address the management and coordination of network, cache, and computation resources in software-defined satellite-terrestrial networks~\cite{Qiu2019Deep}. Zhu \textit{et al.} investigated multi-layer edge collaborative caching in satellite-terrestrial integrated networks to reduce communication latency~\cite{Zhu22Cooperative}. To enhance computational resource utilization and decrease task completion time, it is highly valuable to design a satellite collaborative computing scheme, aiming to achieve the maximum benefits of computation reuse.

To this end, we explore the potential of integrating the computation reuse paradigm with satellite cooperative computing. By leveraging locality-sensitive hashing (LSH) to map preprocessed high-dimensional data, similar input data is grouped into the same hash bucket, which facilitates the reuse of the existing computation results. The similarity between task input data determines how reusable the current task is. Computation reuse is feasible only when data similarity exceeds a specific threshold, enabling input data to leverage processing results from previously cached data and eliminating the need for redundant computations. Building on our implementation of local computation reuse on independent satellites, we propose the concept of \textbf{satellite reuse status (SRS)} to evaluate the current effectiveness of computation reuse on each satellite. A higher SRS value indicates better reuse efficiency and greater potential for collaboration among satellites. Utilizing the SRS metric, we further investigate the feasibility of collaborative computation reuse among satellites and propose a satellite collaborative computation reuse algorithm. \ye{The algorithm establishes an initial collaboration area and expands it when it fails to meet the requirements for inter-satellite collaboration.} This algorithm promotes the sharing of frequently reused entries among high-SRS satellites within these areas, thereby optimizing task completion times and resource utilization through effective computation reuse. 

To the best of our knowledge, this is the first work that integrates the concept of computation reuse into satellite networks to achieve collaborative satellite computing. The key contributions of this paper are summarized as follows:
\begin{itemize}
    \item We investigate the potential of deploying computation reuse in satellite networks and propose a collaborative computation reuse framework \algname for satellite networks. To evaluate the efficiency of computation reuse in satellite systems, we propose the SRS metric, which accounts for real-time reuse rate and computational resource occupancy as key performance indicators.
    

    \item We further explore the feasibility of collaborative computation reuse among satellites and propose a satellite collaborative computation reuse algorithm. We categorize satellites into two types: data source satellites that provide cooperation, and requesting satellites that require assistance. \ye{We expand the collaboration area to maximize the advantages of computation reuse via inter-satellite collaboration}

    \item We conduct extensive simulations using real-world remote sensing datasets to evaluate the effectiveness of the proposed \algname. The simulation results demonstrate that, compared to scenarios without computation reuse and those with local computation reuse on satellites, the proposed algorithm effectively reduces task completion time and computing resource occupancy.
\end{itemize}

The rest of this paper is organized as follows:  Section~\ref{sec:relatedwork} provides an overview of related work. Section~\ref{sec:model} describes our system model with problem formulation. Section~\ref{sec:algorithm} presents the design of our proposed \algname. Section~\ref{sec:evaluation} presents the simulation results to validate the effectiveness of \algname. Section~\ref{sec:conclusion} concludes this paper with future work.

\vspace{0.5cm}
\section{Related Work}~\label{sec:relatedwork}
\vspace{-0.2cm}
\subsection{Computation reuse}
Computation reuse strategies are widely implemented for various computation-intensive processes~\cite{AzadWC22,Barrios2023Service}. For instance, the authors in \cite{guo2018potluck, guo2018foggycache} use computation reuse techniques to mitigate redundant computations in mobile applications and local device caches in nearby servers. Bellal \textit{et al.} introduces a parsing model that integrates computation reuse with task offloading~\cite{bellal2021coxnet}. Drolia \textit{et al.} proposed a system that uses a caching model to adaptively balance the load between the edge and the cloud by leveraging the spatiotemporal locality of requests, offline analysis of applications, and online estimation of network conditions~\cite{Drolia2017Cachier}. Meng \textit{et al.} utilized the similarity of background environment (BE) frames to reduce the bandwidth required for prefetching BE frames from the server by caching and reusing similar frames~\cite{Meng2020Coterie}. Mastorakis \textit{et al.} and Al \textit{et al.} leveraged computation reuse within NDNs, devising an information center network framework and a Reservoir network framework to enhance edge reuse, respectively~\cite{mastorakis2020icedge, al2022reservoir}. Existing work mainly focuses on computation reuse at the edge nodes, which are mainly deployed on resource-abundant devices like multi-access edge computing (MEC) servers at the ground. It is essential to design a computation reuse framework applicable to resource-constraint satellite networks.
\subsection{Satellite collaborative computing}
Satellite communication is expected to provide global coverage to the remote computing services. Transmitting various data via satellites facilitates the sharing of computing resources across diverse geographic areas and the provision of efficient computing services in remote or rural areas without sufficient terrestrial network infrastructure~\cite{ZhangINFOCOM22,ZhuIoT22}. Tang \textit{et al.} proposed a caching strategy between satellites and the ground stations, which utilizes a ridge regression-based regional feature prediction model to divide the LEO satellite coverage area into multiple cooperative areas. A game theory-based cooperative caching algorithm is employed to achieve distributed caching decisions~\cite{tang2024cooperative}. Chen \textit{et al.} proposed a content caching and content distribution method for satellite content delivery networks (CDNs) to solve the problem of rationally deploying CDNs in highly dynamic satellite environments~\cite{Chen2023Cooperative}. Hao \textit{et al.} proposed a regional cooperative caching and distribution strategy to solve the problem of limited satellite resources and the need for a caching strategy to achieve efficient content delivery with less cache redundancy, which divides the network topology into multiple areas and utilizes similar areas to remove redundant content~\cite{Hao2021Co}. Although the previous studies concentrate on satellite-ground and satellite-satellite caching strategies to mitigate communication delays, it is significant to integrate computation reuse with data caching for satellite on-board data processing. In this paper, we design a computation reuse framework aiming at saving computation resource usage and data processing time for both individual satellite computing and inter-satellite computing scenarios.

In summary, the application of computation reuse techniques to collaborative satellite data processing remains a largely unexplored research area. Investigating this domain can further enhance satellite resource utilization, accelerate data processing speeds, and improve the overall quality of satellite services.
\begin{figure}
    \centering
    \includegraphics[width=1\linewidth]{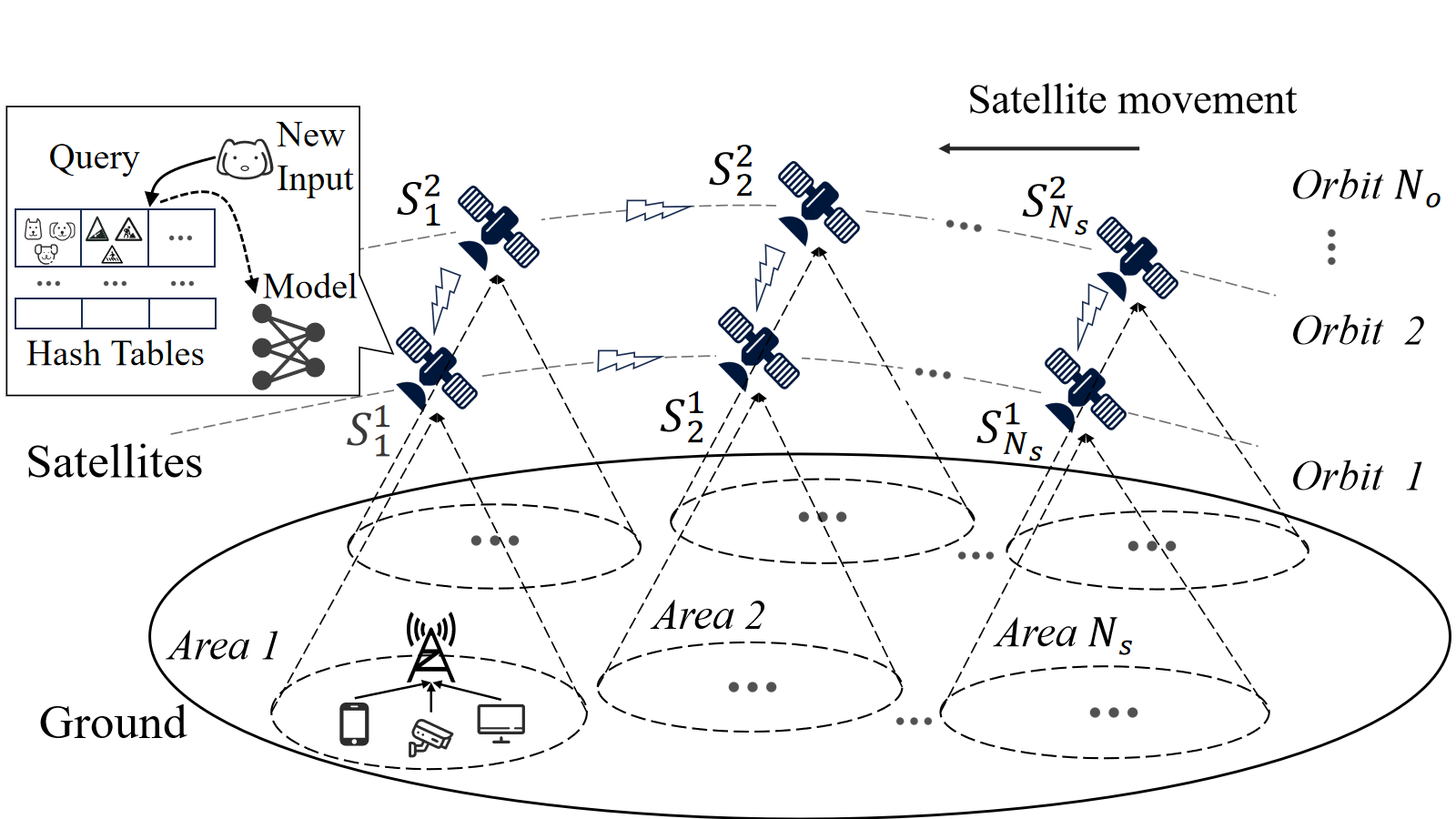}
    \caption{System Model}
    \label{system model}
\end{figure}

\section{System Model}\label{sec:model}

\subsection{Network model}
In this paper, we assume a satellite edge computing network composed of multiple orbits as shown in \figurename~\ref{system model}. The network consists of $N_o$ satellite orbits, each orbit containing $N_s$ satellites, represented as $\mathcal{S}=\{S_1,...,S_{N_s}\}$. Satellites in different orbits are identified by superscripts, i.e., the $n^{th}$ satellite on the $x^{th}$ layer is $S_n^x$. 
Satellites move in the specified direction, with overlapping coverage areas existing between satellites in adjacent orbits as well as those in the same orbit. Additionally, each satellite is equipped with data caching and in-orbit data processing capabilities.




For each satellite, a \textbf{satellite computation reuse table (SCRT)} is used to cache reusable records. The SCRT for satellite $S_n^x$ can be denoted as $SCRT_{S_n^x}$. We assume that each satellite has the same cache size $C^\textit{stg}$. For data on the processing of the arriving tasks $\Gamma^s$ collected by satellite $S_n^x$, we partition it into subtasks $t\in\Gamma^s$. We assume that the satellite server receives and executes the tasks following Little’s Law $M$/$M$/1 queuing system. Then, the reuse record for subtask $t$ can be defined as:
$$record_t=<D_t,P_t,R_t,N_t>$$
where $D_t$ is defined as the input data for executing the target task, $P_t$ is the type of target task to be processed, $R_t$ is the output data after successfully executing the task with the given input, and $N_t$ is the number of times the record has been reused.


During network operation, each satellite collects the data processing tasks corresponding to their respective assigned areas based on the real-world scenario. After receiving the task, the satellite starts in-orbit processing. To minimize resource waste resulting from repetitive computations on highly similar data, each satellite can perform \textit{satellite local computation reuse} by utilizing the similarity of previously processed data. Furthermore, \textit{satellite collaborative computation reuse} is employed to further enhance the quality of task processing.

\subsection{Communication model}




Due to the restrictions imposed by communication distances,
each satellite can only transmit tasks to its adjacent satellites
through ISL~\cite{ChenTVT21}. Assuming the unobstructed line of sight by the Earth, the maximum data rate between satellite $S_k$ and $S_i$ is:

\begin{equation}
r_{S_k,S_i} = B_s \log_2 \left( 1 + \text{SNR}(S_k, S_i) \right)
\end{equation}
where $B_s$ is the channel bandwidth, ${SNR}(S_k, S_i)$ is the signal to noise ratio, which can be
calculated by the following equation~\cite{Leyva2021InterPlane}:

\begin{equation}
\text{SNR}(S_k, S_i) = \frac{Pow_t G_{S_k}(S_i) G_{S_i}(S_k)}{N_0 L(S_k, S_i)}
\end{equation}
where $Pow_t$ is the power of transmitter, $ G_{S_k}(S_i)$ is the average antenna gain of satellite $S_k$ towards satellite $S_i$ and $ G_{S_i}(S_k)$ is the average antenna gain of satellite $S_i$ towards satellite $S_k$. $L(S_k, S_i)$ is the free space path loss calculated by:
\begin{equation}
L(S_k, S_i) = (\frac{4 \pi f_c \, \text{dist}(S_k, S_i)}{c})^2
\end{equation}
where $f_c$ is the carrier frequency, $dist(S_k, S_i)$ is the Euclidean distance between satellite $S_k$ and $S_i$, and $c$ is the speed of the light. $N_0$ is the noise power spectral density calculated by:
\begin{equation}
N_0 = k_B T B_s
\end{equation}
where $k_B$ is the Boltzmann constant, $T$ is the receiver noise temperature. 

Satellite communication consumption primarily stems from inter-satellite collaboration (the sharing of SCRT records among satellites). The size of shared records is primarily determined by their input data $D_t$ and output data $R_t$. When specific conditions are met, the data source satellite $S_i$ distributes high-value records to all satellites in its current collaboration area $\mathcal{S}_{co}=\{S_1,...,S_{co}\}$. More details can be found in the descriptions of the proposed \textit{SCCR} algorithm in Section~\ref{sec:algorithm}.

Therefore, the total communication cost for satellite $S_i$ to complete the arriving task $\Gamma^s$ is:



\begin{equation}
\Psi=\sum_{t\in\Gamma^s} \sum_{S_k \in \mathcal{S}_{co}} \frac{\tau\cdot(D_t+R_t)}{r_{S_k,S_i}}
\end{equation}
where $\tau$ represents the amount of records shared per time.

\subsection{Computation model}

When implementing computation reuse, the satellite first finds the nearest neighbor within the locality sensitive hashing (LSH) table. It then evaluates the similarity between the current task's input and this neighbor to determine whether to reuse the processing results. We assume that the subtask operates under two schemes: execution from scratch on a satellite server, or computation reuse. Let $x_t$ represent a binary variable for computation reuse, where $x_t = 0$ indicates processing the subtask from scratch, and $x_t = 1$ denotes the use of computation reuse. The details of these two schemes are as follows:

\begin{itemize}
    \item \textbf{Computation from scratch}: this operation is defined as:
    \begin{equation}
   \chi_{t}^\text{compute}=(1-x_t)\cdot(W+\frac{F_{t}}{C^\textit{comp}}) 
    \label{eq:scr}
    \end{equation}
     where $W$ denotes the lookup cost (finding the nearest neighbor). When the input data $D_t$ of the current subtask locates its nearest neighbor in the LSH table and their similarity (structural or cosine similarity) exceeds a predefined threshold, $x_t = 1$; otherwise, $x_t = 0$. All subtasks except the first two undergo a lookup operation. Here, $F_t$ represents the  computational resource required to execute subtask $t$, and $C^\textit{comp}$ is the satellite's computational capability.
\end{itemize}

\begin{itemize}
    \item \textbf{Computation with reuse}:
    In this case, if the input data of the subtask is sufficiently similar to the input data of the cached record, the output data of the cached record can be reused, without the need to perform the computation. Instead, only a lookup operation is required as:
    \begin{equation}
    \chi_{t}^\textit{reuse}=x_{t}\cdot W
    \end{equation}
\end{itemize}

For a complete task $\Gamma^s$, the computation cost $\chi$ is calculated as follows:
\begin{align}
    \chi
     &=\sum_{t\in\Gamma^s}{\chi_t^\textit{compute}+\chi}_t^\textit{reuse} \nonumber\\
     &=\sum_{t\in\Gamma^s}{\left[W+(1-x_t)\cdot\frac{F_t}{C^\textit{comp}}\right]}
\end{align}

\subsection{Problem formulation}
The cost required to complete the entire task is defined as:
\begin{equation}
    \varsigma 
    =\alpha\cdot\Psi+\chi
    \label{completiontime}
\end{equation}
where $\alpha$ is a binary parameter for balancing communication consumption with computational consumption.

Based on the above derivation, our goal is to minimize the satellite's task completion time. The optimization problem is defined as follows:
\begin{equation}\label{object function}
    \mathop{\min}\limits_{x_t,\tau}\quad{\varsigma}
\end{equation}

The above optimization objective must satisfy the following constraints: the size of the SCRT stored by each satellite should not exceed the size of its storage space \( C^{\textit{stg}} \), the total number of tasks handled by each satellite should be less than its computational power \( C^{\textit{comp}} \), and the total amount of recorded data transmitted by each satellite when it collaborates should not exceed its communication bandwidth \( B_s \).

\begin{figure}
    \centering
    \includegraphics[width=1\linewidth]{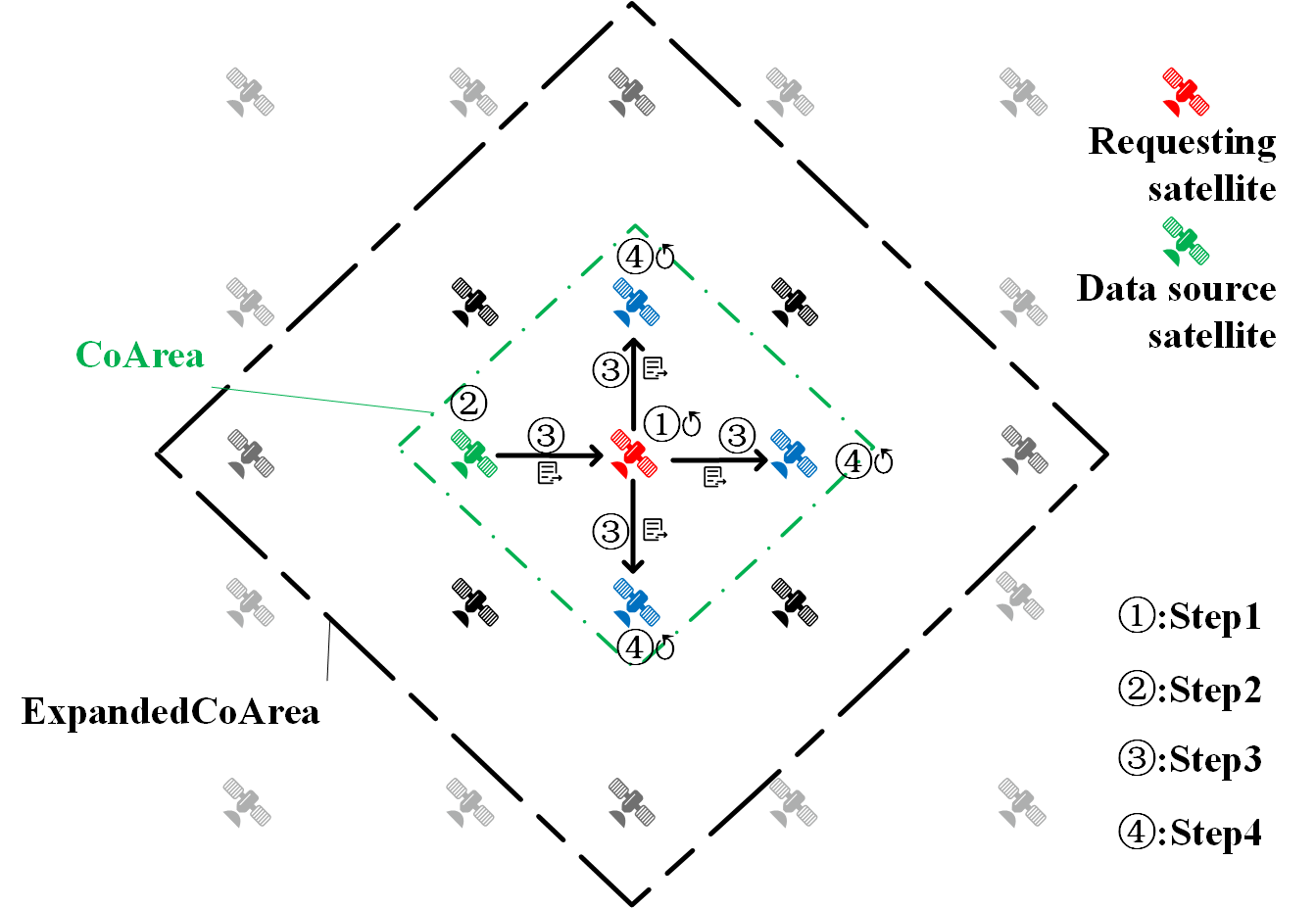}
    \caption{An example of satellite collaborative computation reuse. }
    \label{fig:alg}
\end{figure}

\section{Design of \algname }\label{sec:algorithm} 
\subsection{Overview}
In this section, we present two algorithms based on CCRSat: the \textbf{satellite local computation reuse algorithm (\textit{SLCR})} and the \textbf{satellite collaborative computation reuse algorithm (\textit{SCCR})}. The \textit{SLCR} algorithm elaborates on the entire process of local computation reuse within a satellite, while the \textit{SCCR} algorithm extends this concept by incorporating satellite collaboration to enhance computation reuse efficiency across the satellite network.


To measure the current reuse status of a satellite, we propose the \textbf{Satellite Reuse Status (\textit{SRS})} metric. A higher SRS value signifies greater benefit derived from the computation reuse mechanism. Specifically, the \textit{SRS} is determined by two parameters: the satellite’s current reuse rate and CPU occupancy. Specifically, \textit{SRS} is directly proportional to the reuse rate and inversely proportional to CPU occupancy. Conversely, high CPU occupancy reflects lower reuse gain, as it indicates reliance on pre-trained models for task processing, reducing the efficiency of computation reuse. When executing the \textit{SLCR} algorithm locally, each satellite updates its \textit{SRS} after each computation reuse operation. The \textit{SRS} of satellite $S$ is defined as:
\begin{equation}
\textit{SRS}_{S} = \ \beta \cdot \text{rr}_{S} + (1-\beta) \cdot \left( {1 - {C}_{S}} \right)
\label{SRS}
\end{equation}
where $\text{rr}_{S}$ denotes the reuse rate of $S$, and ${C}_{S}$ denotes the CPU occupancy of $S$. $\beta$ is the weight parameters.



To assess whether a satellite has a favorable reuse status, we define the threshold $th_{co}$. When a satellite's \textit{SRS} drops below $th_{co}$, it indicates that its reuse status is critical, necessitating a request for cooperation from other satellites. Furthermore, it cannot act as a \textbf{data source satellite} ($S_{src}$) for collaborating with other satellites.
 
\figurename~\ref{fig:alg} provides the overview of the collaborative computation reuse process among satellites. This figure shows an example of a satellite requesting cooperation and receiving assistance within the initial collaboration area in a 5$\times$5. The satellite herein is referred to as the \textbf{requesting satellite} ($S_{req}$). The whole process can be divided into four steps (See the arrows in \figurename~\ref{fig:alg}):

\textbf{Step 1}: When a satellite finds that its $\textit{SRS}$ falls below the threshold $th_{co}$, it initiates the collaborative computation reuse process to request support from other satellites within the initial collaboration area.


\textbf{Step 2}: $S_{req}$ retrieves the $\textit{SRS}$ values of all satellites within its initial collaboration  area to identify the satellites with $\textit{SRS}$ values exceeding $th_{co}$ . If such a satellite is identified, the one with the highest $\textit{SRS}$ value is designated as the $S_{src}$ for the collaboration area. If no suitable satellite is found, the collaboration area will be expanded. The details of the collaboration area expansion methodology are discussed below:

The initial collaboration area is constructed by $S_{src}$ and its surrounding satellites. The expanded collaboration area is formed by including the surrounding satellites of all satellites within the initial area.


\textbf{Step 3}: $S_{src}$ broadcasts its top $\tau$ records with the highest reuse count throughout the collaboration area.

\textbf{Step 4}: Each satellite updates its \textbf{SCRT} based on the received records. If a satellite has already cached the records sent by $S_{src}$, no update is needed. When updating, the reuse count is reset to zero to avoid being influenced by the reuse count from $S_{src}$.


\subsection{Satellite local computation reuse}
Algorithm~\ref{alg:OLCR} describes the details of our proposed SLCR algorithm. For the current sub-task \( t \), the inputs are the original input data \( D_t \) and  \( P_t \). This algorithm first pre-processes \( D_t \) to obtain \( PD_t \) (pre-processed input data), which involves resizing, spreading, normalization, and data type conversion for subsequent image processing and model input (line 1). Next, it determines \textbf{SCRT} based on \( P_t \) and hash \( PD_t \) into the corresponding bucket of the SCRT. Then it finds the nearest neighbors of \( PD_t \) in the bucket (line 2). If no candidate can be found, this sub-task cannot be reused. In this case, we use \( P_t \) to select the appropriate pre-trained model to process \( PD_t \), and then form a new record in the SCRT based on \( D_t \), \( P_t \), \( R_t \), and \( N_t \) (initially set to 0) (lines 3-6).

On the other hand, if the nearest neighbor is found, the algorithm calculates the \textbf{structural similarity (SSIM)} between \( PD_t \) and the selected neighbor. SSIM is a metric that quantifies the similarity between two images on a scale from -1 to 1. A value of 1 signifies identical images, 0 indicates no similarity, and negative values denote inverse correlation. The SSIM can be obtained by:
\begin{equation}
\label{eq:ssim}
 \begin{aligned}
&\textrm{SSIM}(x, y) \\&= \left( \frac{2\mu_x \mu_y + C_1}{\mu_x^2 + \mu_y^2 + C_1} \right)
\left( \frac{2\sigma_x\sigma_y + C_2}{\sigma_x^2 + \sigma_y^2 + C_2} \right)
\left( \frac{\sigma_{xy} + C_3}{\sigma_x \sigma_y + C_3} \right)
\end{aligned}
\end{equation}
where $x$ and $y$ represent $PD_t$ and the selected nearest neighbor.

To enhance computation reuse rate, we set a threshold $th_{sim}$ to measure the similarity of the two inputs. Reuse occurs only when $ssim>th_{sim}$. At this point, the algorithm finds the nearest neighbor record and takes its output result as the reuse result $R_t$ for the current calculation, and then increments its reuse count \( N_t \) by 1 (lines 7-11). If $ssim>th_{sim}$ is not satisfied, it processes \( PD_t \) using the pre-trained model, forms a record, and updates it's SCRT accordingly (lines 12-17).

\begin{algorithm}[tb!]
    \caption{SLCR Algorithm}
    \label{alg:OLCR}
    \renewcommand{\algorithmicrequire}{\textbf{Input:}}
    \renewcommand{\algorithmicensure}{\textbf{Output:}}
    \begin{algorithmic}[1]
        \REQUIRE Input data $D_t$, task type $P_t$
        \ENSURE Result of the process $R_t$
        
        \STATE $PD_t$ ← $\textit{Preprocess}(D_t)$;
        \STATE $match$ ← $\textit{FindNearestNeighbor}(P_t,PD_t)$;
        \IF{$match$ = $\emptyset $}
            \STATE $R_t$ ← $\textit{PreTrainedModel}(PD_t,P_t)$;
            \STATE $SCRT$ ← $record=<D_t,P_t,R_t,N_t>$;
            \STATE $LSH(P_t)$ ← $\textit{Renew}(LSH(P_t),PD_t)$;
        \ELSE
            \STATE $ssim$ ← $\textit{SSIM}(I_t,match)$;
            \IF{$ssim > th_{sim}$} \label{th_sim}
                 \STATE $R_t$ ← $\textit{FindOutcome}(match)$;
                 \STATE$ \textit{ReuseCountRenew}(match)$;
            \ELSE
                 \STATE $R_t$ ← $\textit{PreTrainedModel}(PD_t,P_t)$;
                 \STATE $SCRT$ ← $record=<D_t,P_t,R_t,N_t>$;
                 \STATE $LSH(P_t)$ ← $\textit{Renew}(LSH(P_t),PD_t)$;
            \ENDIF
        \ENDIF
        
            
                
            
        
    \end{algorithmic}
\end{algorithm}

\subsection{Satellite collaborative computation reuse}
\begin{algorithm}[tb!]
    \caption{SCCR Algorithm}
    \label{alg: ISCCR}
    \renewcommand{\algorithmicrequire}{\textbf{Input:}}
    \renewcommand{\algorithmicensure}{\textbf{Output:}}
    \begin{algorithmic}[1]
        \REQUIRE collaboration requesting satellite $S_{req}$, satellite network scale $N$
        \ENSURE $S_{src}$, collaboration area $CoArea$
        \STATE $loc_{req}$ ← $\textit{LocateSat}(S_{req})$;
        \STATE $CoArea$ ← $\textit{GetCoArea}(loc_{req},N)$;
        \STATE $S_{max}$ ← $\textit{find\_SRS\_max}(CoArea)$;
        \IF{$S_{max}.SRS > th_{co}$} \label{th_co}
            \STATE $S_{src}$ ← $S_{max}$;
        \ELSE
            \STATE $CoArea$ ← $\textit{GetExpandedCoArea}(CoArea,N)$;
            \STATE $S_{max}$ ← $\textit{find\_SRS\_max}(CoArea)$;
            \IF{$S_{max}.SRS > th_{co}$}
                \STATE $S_{src}$ ← $S_{max}$;
            \ELSE
                \RETURN
            \ENDIF
        \ENDIF
        \RETURN $S_{src}$, $CoArea$
    \end{algorithmic}
\end{algorithm}

The previous Algorithm~\ref{alg:OLCR} focuses on computation reuse within a single satellite. However, in a multi-satellite edge computing network, the data processed by these satellites often exhibit significant similarity. Based on this observation, we propose \textit{SCCR}, an inter-satellite computation reuse algorithm, as outlined in Algorithm~\ref{alg: ISCCR}.


When the $\textit{SRS}$ of satellite (Equation~\ref{SRS}) falls below $th_{co}$, it triggers \textit{SCCR} algorithm and acts as the collaboration request satellite $S_{req}$. \textit{SCCR} algorithm takes $S_{req}$ and the network scale $N$ as inputs. First, the location information of $S_{req}$ is retrieved. Based on the obtained information and $N$, the initial collaboration area \textit{CoArea} is constructed using \textit{GetCoArea()}, which consists of $S_{req}$ and its surrounding satellites. Within this area, the satellite with the highest $\textit{SRS}$ (i.e., the satellite that exhibits the best reuse status in the current collaboration area) is identified, referred to as $S_{max}$ (lines 1-3).

Next, \textit{SCCR} algorithm assess whether $S_{max}$ qualifies as the data source satellite, $S_{src}$. If $S_{max}.SRS > th_{co}$, $S_{max}$ is selected as $S_{src}$ (lines 4-5), otherwise, the collaboration area is expanded to improve the likelihood of finding a suitable satellite. The expanded collaboration area can be constructed using \textit{GetExpandedCoArea()}, which includes all satellites within the initial area, along with their surrounding satellites incorporated into the collaboration scope. Then a search is conducted within this expanded area to find a satellite that satisfies $S_{max}.SRS > th_{co}$. If such a satellite is found, it is designated as $S_{src}$ (lines 6-10). If no suitable satellite is found, the algorithm terminates, concluding the collaboration (lines 11-13). Finally, the algorithm outputs the data source satellite $S_{src}$ and the collaboration area $CoArea$ (lines 14-15). Based on $S_{src}$ and $CoArea$, the top $\tau$ reuse records of $S_{src}$ are shared across the entire collaboration area.


 Regarding the complexity analysis of the proposed algorithm, assuming the network scale is $N\times N$, the time complexity for obtaining both the initial collaboration area and the expanded collaboration area is $O(N^2)$, and the operation of finding the maximum \textit{SRS} is linear with a complexity of $O(N^2)$. Therefore, the overall time complexity is $O(N^2)$. The space complexity is mainly determined by the size of the collaboration area \textit{CoArea}, which in the worst case is $O(N^2)$.

\vspace{0.3cm}
\section{Experimental Evaluation}~\label{sec:evaluation}
\vspace{-0.1cm}

\subsection{Experimental setup}
To evaluate the performance of our proposed \algname, we conduct the experiments within a simulated satellite computing environment. The experimental setup utilizes a PC with an Intel Core i7-10875H processor and 16GB of RAM, operating on Ubuntu 18.04. \algname utilizes the FALCONN library for LSH operations\footnote{\url{https://github.com/FALCONN-LIB/FALCONN}}, specifically using hyperplane hashing. We adopt the UC Merced Land Use dataset~\cite{UC2010Yang}, which is a well-established dataset that includs remote sensing image scenes for land-use classification. We further adjust this dataset to align with the application scenarios of \algname. Specifically, we assume that each satellite cluster processes the same total task volume (625 images with a total size of 12,817 MB), evenly distributed among satellites. The pre-trained model used in the experiments is GoogleNet22~\cite{Szegedy2015Going}. The main experimental parameters are summarized in Table~\ref{tab:param}.

\begin{table}[tb!]
\centering
\caption{Main experimental parameters}\label{tab:param}
\begin{tabular}{lc}
\hline
\textbf{Parameter} & \textbf{Value} \\ \hline 
Network scale ($N \times N$) & $N$ = 5, 7, 9 \\ 
Satellite bandwidth $B_s$ & 20 MHz~\cite{Leyva2021InterPlane}\\
Computational capability $C^{comp}$ & 3~GHz~\cite{zhang2023satellite}\\
Number of hash tables $p_l$ & 1 \\ 
Number of hash functions $p_k$ & 2 \\ 
Weight parameters $\beta$ & 0.5\\
Input similarity threshold $th_{sim}$ & 0.7 \\
Number of satellite broadcast records $\tau$ & 11 (default) \\ 
Cooperation request threshold ${th_{co}}$ & 0.5 (default) \\ \hline
\end{tabular}
\end{table}

We use the following criteria to evaluate the performance of \algname in the experiments: \begin{enumerate}
    \item  \textbf{Task Completion Time} is the total time taken for all satellites within the network to process the respective tasks and obtain results. This encompasses both task execution from scratch and instances where computation reuse (either local or collaborative) is applied; 
\item  \textbf{Reuse Rate} is the average proportion of reused tasks relative to the total number of tasks when (collaborative) computation reuse is implemented within the satellite network;
\item  \textbf{CPU Occupancy} is the average CPU occupancy of satellites from task receipt to task completion; 
\item  \textbf{Reuse Accuracy} is the ratio of correctly reused tasks to total reused tasks;
\item  \textbf{Data Transfer Volume} is the total data transfer volume of all satellites in the entire network.

\end{enumerate}

To ensure reproducibility and generalizability, all experiments are conducted across satellite networks on two different scales. The experimental scenarios comprised the following categories: \begin{itemize}
    \item \textbf{w/o CR} does not apply computation reuse, i.e., satellites process each task from scratch.
    \item \textbf{SRS Priority} searches for satellites with optimal SRS values within the entire satellite network and collaborates, the broadcast area is also the entire satellite network. 
    \item \textbf{SLCR} only applies local computation reuse, i.e., each satellite reuses computations locally (Algorithm~\ref{alg:OLCR} in Section~\ref{sec:algorithm}).
    \item \textbf{SCCR-INIT} is an initial version of our proposed satellite collaborative computation reuse algorithm that does not include expansion of the collaboration area.
    \item \textbf{SCCR} is our proposed satellite collaborative computation reuse algorithm (Algorithm~\ref{alg: ISCCR} in Section~\ref{sec:algorithm}).
\end{itemize}

Furthermore, a sensitivity analysis is performed on critical parameters that could have a substantial impact on the experimental results. These parameters include:\begin{itemize}
    \item \textbf{$\tau$} is the number of locality-sensitive hashing records that the source satellite broadcasts.
    \item \textbf{$th_{co}$} is a threshold used to determine the necessity of collaboration (See line~\ref{th_co} in Algorithm~\ref{alg: ISCCR}); if the value of \textit{SRS} is below this threshold, it means that local reuse is suboptimal or remains suboptimal even after collaborative computation reuse.
\end{itemize}  

\subsection{Task processing performance}



\begin{table}[tb!]
\caption{Reuse accuracy for different scenarios.}\label{tab:accuracy}
\scalebox{1}{
\begin{tabular}{c|cc|ccc}
\hline
NW Scale & w/o CR & SRS Priority & SLCR & SCCR-INIT & SCCR \\ \hline
5$\times$5     & 1       & 0.9692             & 1     & 0.9980          & 0.9970     \\
7$\times$7     & 1      & 0.9756             & 1     & 0.9974          & 0.9954     \\
9$\times$9     & 1       & 0.9190             & 1     & 0.9757          & 0.9750     \\ \hline
\end{tabular}
}
\end{table}

\begin{figure}[t]
	\centering
	\begin{minipage}[b]{.85\columnwidth}
		\centering
		\includegraphics[width=\columnwidth]{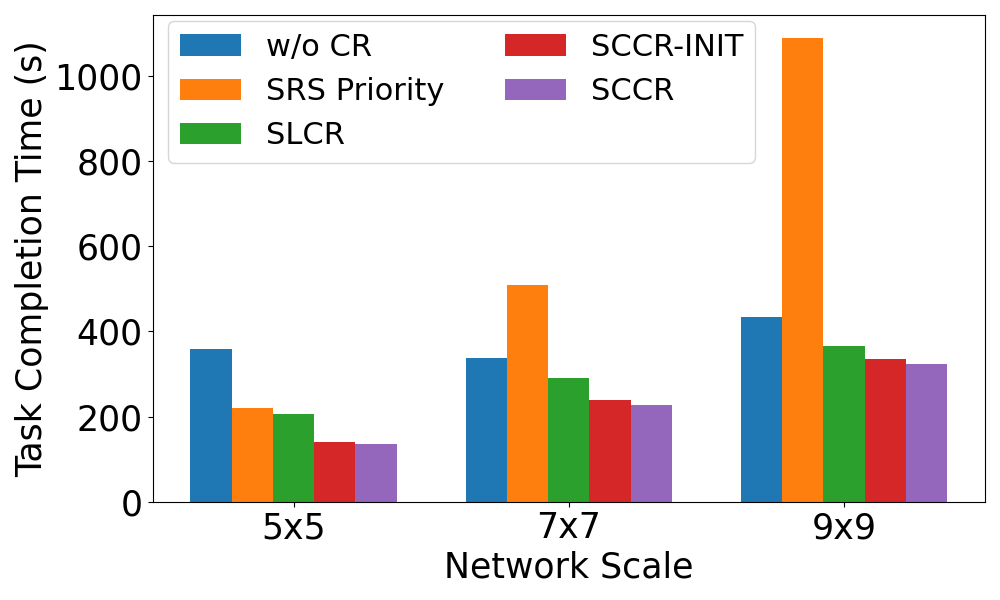}
		\subcaption{Task completion time }\label{fig:1_time}
        \vspace{0.4cm}
	\end{minipage}
        \begin{minipage}[b]{.85\columnwidth}
		\centering
		
		\includegraphics[width=\columnwidth]{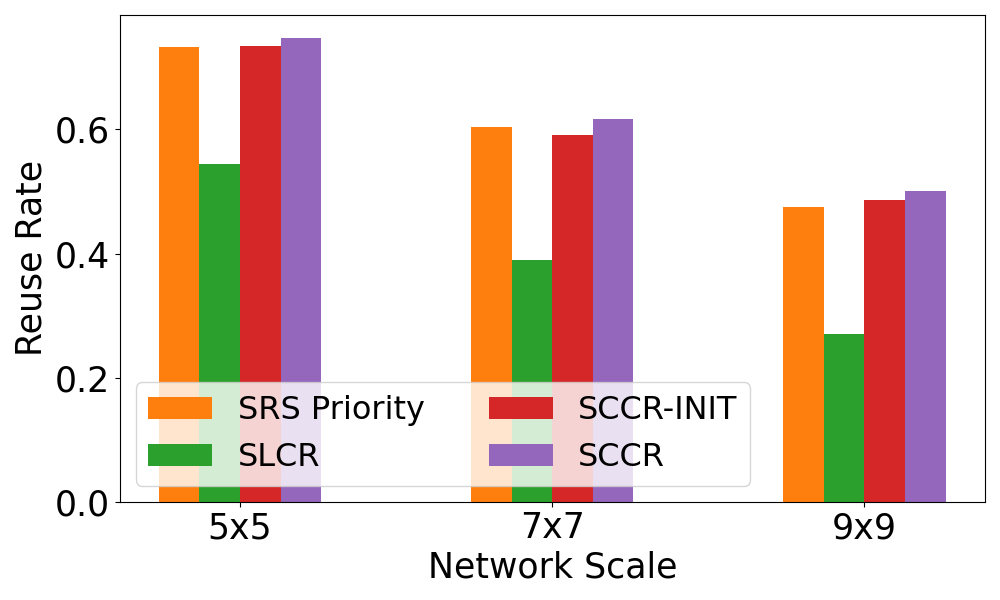}
		\subcaption{Reuse rate}\label{fig:2_rr}
        \vspace{0.4cm}
	\end{minipage}
	\begin{minipage}[b]{.85\columnwidth}
		\centering
		\includegraphics[width=\columnwidth]{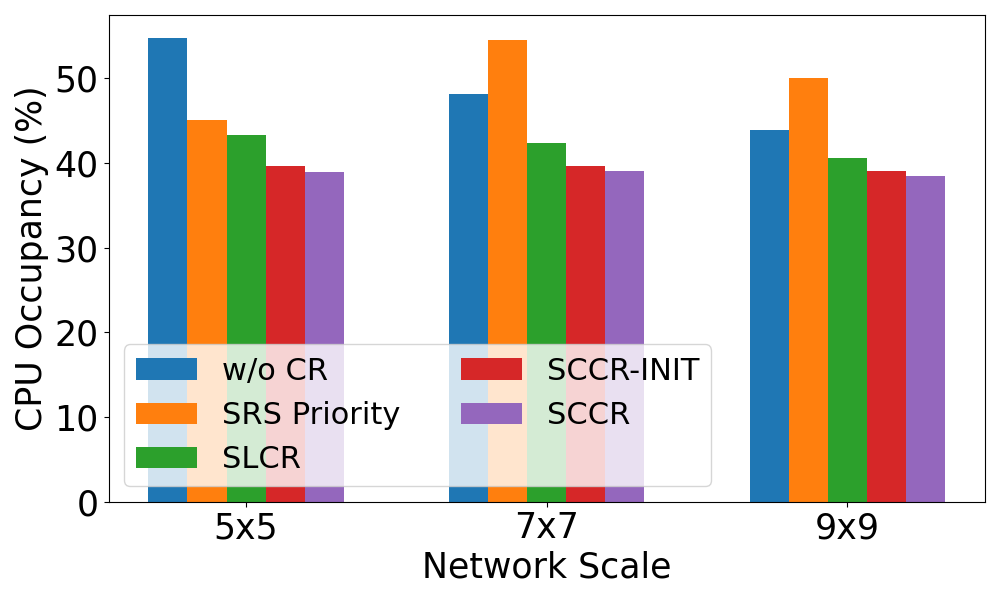}
		\subcaption{CPU occupancy}\label{fig:3_cpu}
	\end{minipage}
    
	\caption{Task processing performance for different scenarios.}
	\label{fig:loss}
   
\end{figure}

Table~\ref{tab:accuracy} summarizes the reuse accuracy achieved by different scenarios. As observed in the table, the reuse accuracy of \textit{SCCR-INIT} and \textit{SCCR} 
is lower compared to \textit{SLCR}, primarily due to the following two key factors:\begin{itemize}
    \item Data correlation: In smaller collaboration areas, satellites process highly correlated data. As the area expands to encompass satellites from diverse orbits and tasks, the heterogeneity of data grows, resulting in a rise in reuse errors among satellites. 
    \item Accumulated errors: In small-scale satellite networks, errors tend to remain confined to specific areas. However, in large-scale networks, computation results propagate across multiple areas, leading to error amplification and decline in accuracy.
\end{itemize}

To mitigate the impact of local computation reuse errors on collaborative computation reuse accuracy, it is crucial to fine-tuning the parameter $th_{sim}$ (line~\ref{th_sim} in Algorithm~\ref{alg:OLCR}). Based on our extensive evaluation, we choose $th_{sim}=0.7$ because it allows \textit{SLCR} to consistently achieve 100\% reuse accuracy across all network scales at this threshold value. For this reason, despite the decline in reuse accuracy, the achieved reuse accuracy of \textit{SCCR} remains satisfactory and surpasses that of \textit{SRS Priority}, which searches for satellites with optimal SRS values across the network.

\figurename~\ref{fig:loss} shows the task processing performance obtained by different scenarios in terms of task completion time, reuse rate, and CPU occupancy. The results indicate that the proposed \textit{SCCR} consistently outperforms other scenarios across all criteria and network scales. In smaller networks, each satellite handles a larger workload, leading to higher data redundancy and reuse rates. The reuse rates for \textit{SLCR} across three network sizes are $0.544$, $0.39$, and $0.27$, respectively. This indicates that smaller networks achieve higher reuse rates, reducing task completion time.

\begin{table}[tb!]
\caption{Data transfer volume (MB) for different scenarios.}\label{tab:bandwidth}
\scalebox{1}{
\begin{tabular}{c|cc|ccc}
\hline
NW Scale & w/o CR & SRS Priority & SLCR & SCCR-INIT & SCCR \\ \hline
5$\times$5     & 0       & 8114.67             & 0     & 889.98          & 1054.09     \\
7$\times$7     & 0       & 44070.41             & 0     & 1732.42          & 1743.56     \\
9$\times$9     & 0       & 184587.78             & 0     & 3125.06          & 3369.23     \\ \hline
\end{tabular}
}
\end{table}

In 5$\times$5 satellite network, \textit{SCCR} reduces task completion time by 62.1\% compared to \textit{w/o CR} (\figurename~\ref{fig:1_time}) and decreases CPU utilization by 28.8\%. Compared to \textit{SLCR}, \textit{SCCR} improves the reuse rate by 37.3\% (\figurename~\ref{fig:2_rr}). This improvement stems from increased task similarity across satellites. \textit{SLCR} relies solely on local computation history, limiting its ability to leverage similar tasks processed by neighboring satellites. By introducing collaboration and using \textit{SRS} to assess reuse status, \textit{SCCR} enables controlled cooperation. This allows satellites with higher reuse potential to share their results efficiently, improving overall reuse rates. Meanwhile, \textit{SCCR} outperforms \textit{SCCR-INIT} due to its gradual expansion of the collaboration area and on-demand collaboration requests. These strategies reduce redundant cooperation, preventing unnecessary delays and excessive data transfer. Naturally, expanding the collaboration area increases total data transmission. As summarized in Table~\ref{tab:bandwidth}, \textit{SCCR-INIT} results in 
889.98~\textit{MB} of data transfer, whereas \textit{SCCR} increases this volume by 18.44\%. In contrast, \textit{SRS Priority} lacks a gradual expansion mechanism, leading source satellites to share records across the entire network. As a result, it performs poorly in terms of task completion time, CPU utilization, and data transfer, with its data transfer volume being 7.7 times higher than that of \textit{SCCR}. In 7$\times$7 network, its task completion time exceeds \textit{w/o CR} by 41.1\%.

Regarding the results of CPU utilization across different network sizes, \figurename~\ref{fig:3_cpu} shows the trends consistent with task completion time. Since \textit{w/o CR} does not leverage computation reuse, it has the highest CPU occupancy. \textit{SCCR} achieves the best performance by maximizing reuse across the network, reducing overall computational workload. Through reuse, resource-intensive tasks are replaced by previously computed results, transforming execution from costly computation to efficient lookup and similarity assessment.

For larger-scale satellite networks, such as 7$\times$7 and 9$\times$9, the performance trends align with those observed in 5$\times$5 configuration, i.e., \textit{SCCR} outperforms other comparative scenarios. Specifically, in 7$\times$7 network, \textit{SCCR} increases data transfer by just 11.14 \textit{MB} compared to \textit{SCCR-INIT}, while reducing task completion time by 4.8\%, highlighting its scalability.

\begin{figure}[tb!]
	\centering
	\includegraphics[width=0.85\linewidth]{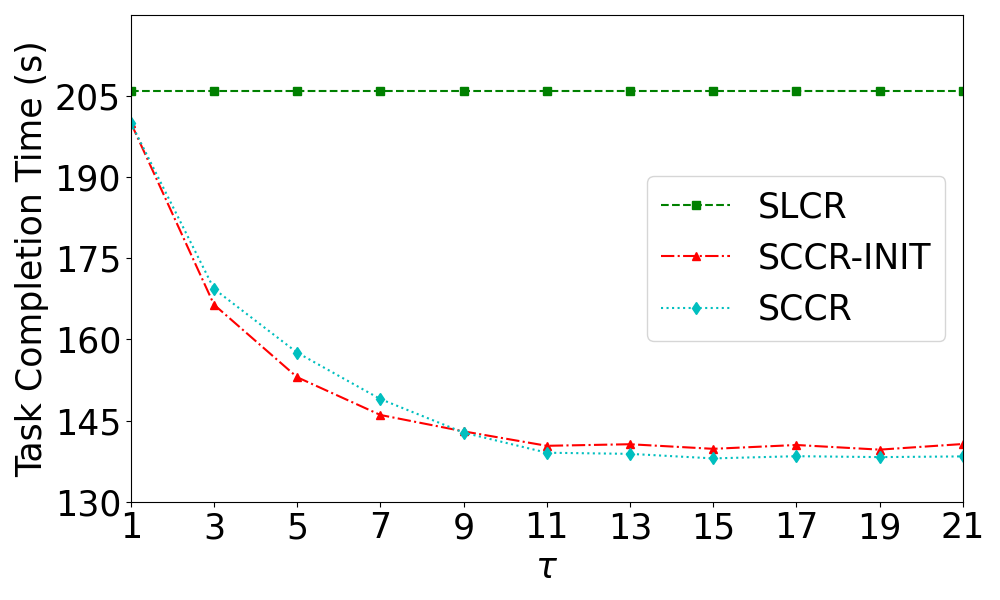}
        \caption{Impact of parameter $\tau$ on task completion time.}
       
    \label{fig:third_row}
\end{figure}

\begin{figure}[tb!]
	\centering
	\includegraphics[width=0.85\linewidth]{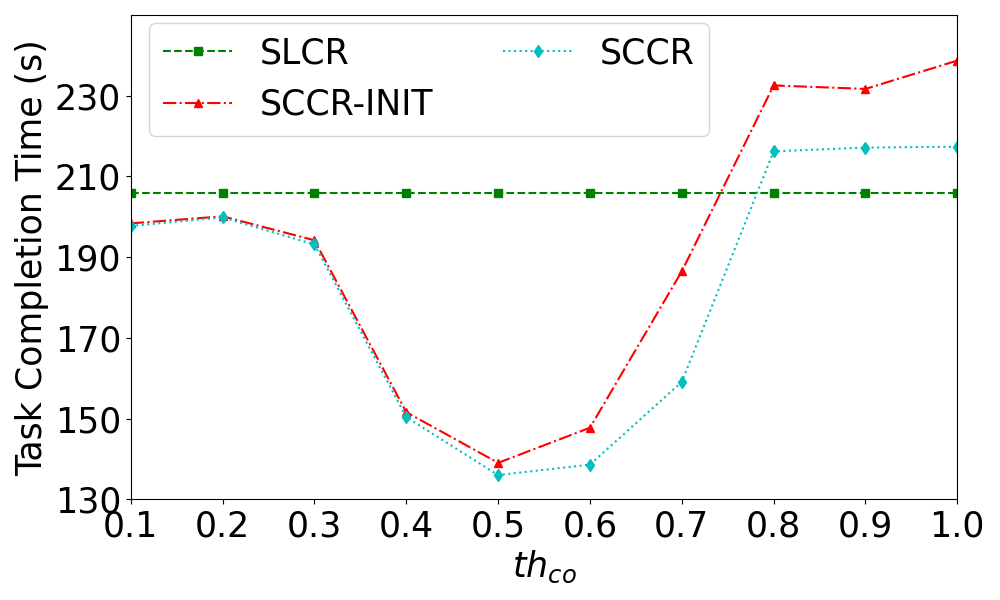}
         \caption{Impact of parameter $th_{co}$ on task completion time.}
	\label{fig:fourth_row}
		

\end{figure}

\subsection{Sensitive analysis}
To assess the impact of key parameters on the performance of \textit{SCCR}, we conduct a sensitivity analysis focusing on two critical parameters $\tau$ and $th_{co}$ within 5$\times$5 network:
\begin{enumerate}
    \item  \textbf{Impact of $\tau$}: It is a parameter that regulates the number of shared records. \figurename~\ref{fig:third_row} illustrates the impact of $\tau$ on task completion time. The task completion time of the \textit{SCCR} algorithm decreases as $\tau$ increases in both the non-extended and extended collaboration area. It stabilizes when $\tau = 11$. A larger $\tau$ allows high-value records to propagate and converge more quickly across the collaboration area. This improves the satellite reuse rate within the network. The results eventually stabilize due to the limited storage capacity of the  satellites. Once the SCRT's storage limit is reached, increasing $\tau$ further provides no additional benefits. Compared to \textit{SCCR-INIT}, \textit{SCCR} achieves better performance. The extended collaboration area enables collaboration when the surrounding satellites cannot support the current satellite.
    
    
    \item  \textbf{Impact of threshold $th_{co}$}: As shown in \figurename~\ref{fig:fourth_row}, \textit{SCCR-INIT} and \textit{SCCR} exhibit similar trends. Task completion time initially decreases as  $th_{co}$ increases but then increases as  $th_{co}$ continues to grow. Both algorithms achieve optimal performance at $th_{co}=0.5$. A very small $th_{co}$ restricts collaboration requests among satellites. Conversely, an excessively large $th_{co}$ leads to unnecessary cooperation. This imposes a heavy communication burden on satellites. When $th_{co}>0.8$, excessive collaboration overhead causes \textit{SCCR} to perform worse than \textit{SLCR}.
\end{enumerate}

\subsection{Summary}
The proposed \textit{SCCR} scenario effectively enhances task processing efficiency in satellite networks by leveraging satellite collaborative computation reuse. Compared to existing scenarios, it achieves significant reductions in task completion time and CPU utilization while maintaining a high reuse rate. The progressive expansion strategy minimizes redundant cooperation and balances performance with data transfer volume. Experimental results confirm that \textit{SCCR} scales well with increasing network size, making it a promising solution for computation reuse in large-scale satellite networks.

\vspace{0.5cm}
\section{Conclusion}~\label{sec:conclusion}
In this paper, we present CCRSat, a framework designed to enable computation reuse in satellite edge computing environments. Based on this framework, we develop a local computation reuse algorithm, which significantly reduces the demand on satellite computing resources and greatly decreases task completion times across the satellite network. Furthermore, we propose a collaborative computation reuse algorithm to utilize inter-satellite collaboration. The experimental results demonstrate that our proposed \algname effectively exploits the potential of computation reuse within the satellite network and significantly outperforms the local computation reuse scenario and cases without computation reuse in terms of various criteria, including task completion time and computational resource occupancy. In our future work, we will explore the use of artificial intelligence (AI)-based techniques to predict SCRT records, aiming to further improve data processing performance in large-scale satellite networks.

\vspace{0.5cm}

\bibliographystyle{ieeetr} 
\bibliography{ref}

\begin{thebibliography}{10}

\bibitem{CentenaroCST21}
M.~Centenaro, C.~E. Costa, F.~Granelli, C.~Sacchi, and L.~Vangelista, ``A survey on technologies, standards and open challenges in satellite {IoT},'' {\em IEEE Communications Surveys \& Tutorials}, vol.~23, no.~3, pp.~1693--1720, 2021.

\bibitem{KodheliCST21}
O.~Kodheli, E.~Lagunas, N.~Maturo, S.~K. Sharma, B.~Shankar, J.~F.~M. Montoya, J.~C.~M. Duncan, D.~Spano, S.~Chatzinotas, S.~Kisseleff, J.~Querol, L.~Lei, T.~X. Vu, and G.~Goussetis, ``Satellite communications in the new space era: A survey and future challenges,'' {\em IEEE Communications Surveys \& Tutorials}, vol.~23, no.~1, pp.~70--109, 2021.

\bibitem{Al-HraishawiCST23}
H.~Al-Hraishawi, H.~Chougrani, S.~Kisseleff, E.~Lagunas, and S.~Chatzinotas, ``A survey on nongeostationary satellite systems: The communication perspective,'' {\em IEEE Communications Surveys \& Tutorials}, vol.~25, no.~1, pp.~101--132, 2023.

\bibitem{ShenCSUR23}
Z.~Shen, J.~Jin, C.~Tan, A.~Tagami, S.~Wang, Q.~Li, Q.~Zheng, and J.~Yuan, ``A survey of next-generation computing technologies in space-air-ground integrated networks,'' {\em ACM Computing Surveys}, vol.~56, no.~1, 2023.

\bibitem{WangIoT23}
S.~Wang and Q.~Li, ``Satellite computing: Vision and challenges,'' {\em IEEE Internet of Things Journal}, vol.~10, no.~24, pp.~22514--22529, 2023.

\bibitem{He2017Exp}
X.~He, S.~Jiang, W.~Lu, G.~Yan, Y.~Han, and X.~Li, ``Exploiting the potential of computation reuse through approximate computing,'' {\em IEEE Transactions on Multi-Scale Computing Systems}, vol.~3, no.~3, pp.~152--165, 2017.

\bibitem{NourNW20}
B.~Nour, S.~Mastorakis, and A.~Mtibaa, ``Compute-less networking: Perspectives, challenges, and opportunities,'' {\em IEEE Network}, vol.~34, no.~6, pp.~259--265, 2020.

\bibitem{Barrios2023Service}
C.~Barrios and M.~Kumar, ``Service caching and computation reuse strategies at the edge: A survey,'' {\em ACM Computing Surveys}, vol.~56, no.~2, 2023.

\bibitem{guo2018foggycache}
P.~Guo, B.~Hu, R.~Li, and W.~Hu, ``Foggycache: Cross-device approximate computation reuse,'' in {\em Proceedings of the international conference on mobile computing and networking (MobiCom)}, pp.~19--34, 2018.

\bibitem{guo2018potluck}
P.~Guo and W.~Hu, ``Potluck: Cross-application approximate deduplication for computation-intensive mobile applications,'' in {\em Proceedings of the International Conference on Architectural Support for Programming Languages and Operating Systems (ASPLOS)}, pp.~271--284, 2018.

\bibitem{mastorakis2020icedge}
S.~Mastorakis, A.~Mtibaa, J.~Lee, and S.~Misra, ``Icedge: When edge computing meets information-centric networking,'' {\em IEEE Internet of Things Journal}, vol.~7, no.~5, pp.~4203--4217, 2020.

\bibitem{al2022reservoir}
M.~W. Al~Azad and S.~Mastorakis, ``Reservoir: Named data for pervasive computation reuse at the network edge,'' in {\em Proceedings of the IEEE International Conference on Pervasive Computing and Communications (PerCom)}, pp.~141--151, 2022.

\bibitem{Nour2022Federated}
B.~Nour, S.~Cherkaoui, and Z.~Mlika, ``Federated learning and proactive computation reuse at the edge of smart homes,'' {\em IEEE Transactions on Network Science and Engineering}, vol.~9, no.~5, pp.~3045--3056, 2022.

\bibitem{ChenTIV23}
L.~Chen, Y.~Li, C.~Huang, B.~Li, Y.~Xing, D.~Tian, L.~Li, Z.~Hu, X.~Na, Z.~Li, S.~Teng, C.~Lv, J.~Wang, D.~Cao, N.~Zheng, and F.-Y. Wang, ``Milestones in autonomous driving and intelligent vehicles: Survey of surveys,'' {\em IEEE Transactions on Intelligent Vehicles}, vol.~8, no.~2, pp.~1046--1056, 2023.

\bibitem{YushanCST21}
Y.~Siriwardhana, P.~Porambage, M.~Liyanage, and M.~Ylianttila, ``A survey on mobile augmented reality with {5G} mobile edge computing: Architectures, applications, and technical aspects,'' {\em IEEE Communications Surveys \& Tutorials}, vol.~23, no.~2, pp.~1160--1192, 2021.

\bibitem{Leyva-MayorgaTC23}
I.~Leyva-Mayorga {\em et~al.}, ``Satellite edge computing for real-time and very-high resolution earth observation,'' {\em IEEE Transactions on Communications}, vol.~71, no.~10, pp.~6180--6194, 2023.

\bibitem{GaoIoT24}
X.~Gao, Y.~Hu, Y.~Shao, H.~Zhang, Y.~Liu, R.~Liu, and J.~Zhang, ``Hierarchical dynamic resource allocation for computation offloading in {LEO} satellite networks,'' {\em IEEE Internet of Things Journal}, vol.~11, no.~11, pp.~19470--19484, 2024.

\bibitem{Hao2023Joint}
Y.~Hao, Z.~Song, Z.~Zheng, Q.~Zhang, and Z.~Miao, ``Joint communication, computing, and caching resource allocation in {LEO} satellite {MEC} networks,'' {\em IEEE Access}, vol.~11, pp.~6708--6716, 2023.

\bibitem{Qiu2019Deep}
C.~Qiu, H.~Yao, F.~R. Yu, F.~Xu, and C.~Zhao, ``Deep {Q}-learning aided networking, caching, and computing resources allocation in software-defined satellite-terrestrial networks,'' {\em IEEE Transactions on Vehicular Technology}, vol.~68, no.~6, pp.~5871--5883, 2019.

\bibitem{Zhu22Cooperative}
X.~Zhu, C.~Jiang, L.~Kuang, and Z.~Zhao, ``Cooperative multilayer edge caching in integrated satellite-terrestrial networks,'' {\em IEEE Transactions on Wireless Communications}, vol.~21, no.~5, pp.~2924--2937, 2022.

\bibitem{AzadWC22}
M.~W. Al~Azad and S.~Mastorakis, ``The promise and challenges of computation deduplication and reuse at the network edge,'' {\em IEEE Wireless Communications}, vol.~29, no.~6, pp.~112--118, 2022.

\bibitem{bellal2021coxnet}
Z.~Bellal, B.~Nour, and S.~Mastorakis, ``Coxnet: A computation reuse architecture at the edge,'' {\em IEEE transactions on green communications and networking}, vol.~5, no.~2, pp.~765--777, 2021.

\bibitem{Drolia2017Cachier}
U.~Drolia, K.~Guo, J.~Tan, R.~Gandhi, and P.~Narasimhan, ``Cachier: Edge-caching for recognition applications,'' in {\em Proceedings of the IEEE International Conference on Distributed Computing Systems (ICDCS)}, pp.~276--286, 2017.

\bibitem{Meng2020Coterie}
J.~Meng, S.~Paul, and Y.~C. Hu, ``Coterie: Exploiting frame similarity to enable high-quality multiplayer {VR} on commodity mobile devices,'' in {\em Proceedings of the International Conference on Architectural Support for Programming Languages and Operating Systems (ASPLOS)}, p.~923–937, 2020.

\bibitem{ZhangINFOCOM22}
Y.~Zhang, Q.~Wu, Z.~Lai, and H.~Li, ``Enabling low-latency-capable satellite-ground topology for emerging {LEO} satellite networks,'' in {\em Proceedings of the IEEE Conference on Computer Communications (INFOCOM)}, pp.~1329--1338, 2022.

\bibitem{ZhuIoT22}
X.~Zhu and C.~Jiang, ``Integrated satellite-terrestrial networks toward {6G}: Architectures, applications, and challenges,'' {\em IEEE Internet of Things Journal}, vol.~9, no.~1, pp.~437--461, 2022.

\bibitem{tang2024cooperative}
J.~Tang, J.~Li, X.~Chen, K.~Xue, L.~Zhang, Q.~Sun, and J.~Lu, ``Cooperative caching in satellite-terrestrial integrated networks: A region features aware approach,'' {\em IEEE Transactions on Vehicular Technology}, 2024.

\bibitem{Chen2023Cooperative}
Y.~Chen, X.~Ma, A.~Zhou, and S.~Wang, ``Cooperative content caching and distribution for satellite {CDNs},'' in {\em Proceedings of the IEEE International Conference on Network Protocols (ICNP)}, pp.~1--6, 2023.

\bibitem{Hao2021Co}
L.~Hao, P.~Ren, and Q.~Du, ``Cooperative regional caching and distribution in space-terrestrial integrated networks,'' in {\em Proceedings of the IEEE/CIC International Conference on Communications in China (ICCC)}, pp.~1042--1047, 2021.

\bibitem{ChenTVT21}
Q.~Chen, G.~Giambene, L.~Yang, C.~Fan, and X.~Chen, ``Analysis of inter-satellite link paths for {LEO} mega-constellation networks,'' {\em IEEE Transactions on Vehicular Technology}, vol.~70, no.~3, pp.~2743--2755, 2021.

\bibitem{Leyva2021InterPlane}
I.~Leyva-Mayorga, B.~Soret, and P.~Popovski, ``Inter-plane inter-satellite connectivity in dense {LEO} constellations,'' {\em IEEE Transactions on Wireless Communications}, vol.~20, no.~6, pp.~3430--3443, 2021.

\bibitem{UC2010Yang}
Y.~Yang and S.~Newsam, ``Bag-of-visual-words and spatial extensions for land-use classification,'' in {\em Proceedings of the SIGSPATIAL International Conference on Advances in Geographic Information Systems (GIS)}, p.~270–279, 2010.

\bibitem{Szegedy2015Going}
C.~Szegedy, W.~Liu, Y.~Jia, P.~Sermanet, S.~Reed, D.~Anguelov, D.~Erhan, V.~Vanhoucke, and A.~Rabinovich, ``Going deeper with convolutions,'' in {\em Proceedings of the IEEE Conference on Computer Vision and Pattern Recognition (CVPR)}, pp.~1--9, 2015.

\bibitem{zhang2023satellite}
H.~Zhang, R.~Liu, A.~Kaushik, and X.~Gao, ``Satellite edge computing with collaborative computation offloading: An intelligent deep deterministic policy gradient approach,'' {\em IEEE Internet of Things Journal}, vol.~10, no.~10, pp.~9092--9107, 2023.

\end{thebibliography}

\end{document}